# No Universal Hyperbola: A Formal Disproof of the Epistemic Trade-Off Between Certainty and Scope in Symbolic and Generative AI

*Generoso Immediato*

generoso.immediato@gmail.com

***Abstract***

*In direct response to requests for a logico-mathematical test of the conjecture, we formally disprove a recently conjectured artificial intelligence trade-off between epistemic certainty and scope in its published universal hyperbolic product form, as introduced in Philosophy & Technology. Certainty is defined as the worst-case correctness probability over the input space, and scope as the sum of the Kolmogorov complexities of the input and output sets. Using standard facts from coding theory and algorithmic information theory, we show, first, that when the conjecture is instantiated with prefix (self-delimiting, prefix-free) Kolmogorov complexity, it leads to an internal inconsistency, and second, that when it is instantiated with plain Kolmogorov complexity, it is refuted by a constructive counterexample. These results establish a main theorem: contrary to the conjecture's claim, no universal "certainty-scope" hyperbola holds as a general bound under the published definitions. We further show that a subsequent "entropy-based" revision—replacing Kolmogorov scope with Shannon joint entropy and redefining the epistemic certainty level accordingly—cannot restore universality either.*

---

## 1. Introduction

- **Motivation**. In response to the explicit invitation in Floridi (2025a, 2025b), we provide a formal logical–mathematical test of the conjecture. The correction paper (Floridi, 2025a) reaffirms the conjecture in its product form $C(M) \cdot S(M) \leq k$ with $k \in (0,1)$, and it retains the Kolmogorov-based definition of scope. Our disproof therefore targets the conjecture in its corrected—and hence most charitable—published formulation. We introduce only standard formal conventions required to render the stated primitives precise and testable (§2.1).

- **Technical result**. We find that under the published definitions, a trade-off between epistemic certainty and scope—specifically, a universal hyperbolic product bound—cannot be satisfied by all artificial intelligence (AI) mechanisms.
- **State of the Art.** Since the conjecture has also prompted a published revision proposal that replaces Kolmogorov-based scope with Shannon joint entropy (and revises certainty accordingly), we add Appendix A to show that this modification still cannot yield a universal constant: unless protocols are explicitly restricted, the product form remains unbounded across admissible mechanisms.

### 1.1. Organization

- Citing the original conjecture's primitives and definition (§2) for rigor and exposition clarity of the core topics.
- Fixing the minimal Algorithmic Information Theory (AIT) conventions required to make the conjecture claim testable (§2.1).
- Presenting the formal disproof in two independent branches—the Prefix-K *inconsistency* (§3.1) and the Plain-K *existential counterexample* (§3.2)—followed by the combined *main theorem* (§3.3) and related robustness notes (§4).
- Reporting the immediate implications of the formal disproof (§5) and conclusions (§6).
- Addressing a published entropy-based scope revision (Shannon joint entropy) and showing that it likewise cannot restore universality (Appendix A).

## 2. Conjecture's Definitions

We reproduce the conjecture and its associated quantities exactly as delineated in the original publication, restricting to the minimal technical information required for this formal rebuttal. Additional brief notes make explicit the conventions needed for mathematical testability; for the full conceptual motivation, see Floridi (2025a, 2025b).

I. **Conjecture's claim:** Let $\mathsf{M}$ be an AI mechanism modeled as $f_{\mathsf{M}}\colon \mathsf{I} \to \mathsf{O}$ over non-empty finite sets $\mathsf{I}$ (input) and $\mathsf{O}$ (output). The conjecture posits a general inverse relation between certainty $\mathsf{C(M)}$ and scope $\mathsf{S(M)}$: $\exists \mathsf{k} \in (0, 1)$, independent of $\mathsf{M}$, such that $\mathsf{C(M)} \cdot \mathsf{S(M)} \leq \mathsf{k}$, for all 'sufficiently expressive' $\mathsf{M}$.

   Immediate corollaries as stated in the original conjecture:

   - **Provable-certainty regime** (corollary)**:** If $\mathsf{C(M)} = 1 \Rightarrow \mathsf{S(M)} \leq \mathsf{k}$
   - **Wide-scope regime** (corollary)**:** If $\mathsf{S(M)} > \mathsf{k} \Rightarrow \mathsf{C(M)} < 1$

For effectiveness and clarity, we restate only the scalar measures $\mathsf{C}(\mathsf{M})$ and $\mathsf{S}(\mathsf{M})$, omitting the full explanation and interpretive discussion that accompany their definitions.

II. **Epistemic Certainty**: $\mathsf{C}(\mathsf{M}) \coloneqq 1 - \sup_{x \in \mathrm{I}} \Pr[f_M(x) \neq \mathsf{Spec}(x)]$

**Note 2.1:** For each fixed $x \in \mathrm{I}$, $f_{\mathrm{M}}(x)$ is treated as an O-valued random variable; thus, $\Pr[f_M(x) \neq \mathsf{Spec}(x)]$ denotes the error probability at that fixed input (degenerating to an indicator in the deterministic case). The published definition does not specify the underlying source of randomness; accordingly, we take $\Pr[\cdot]$ with respect to whatever stochasticity or uncertainty is operative (e.g., internal randomization, nondeterministic execution, or exogenous perturbations), without committing to any particular noise model. This convention is the minimal probabilistic reading consistent with the published primitives for $\mathsf{C}(\mathsf{M})$, namely an error probability evaluated at a fixed x and then taken in the worst case over $x \in \mathrm{I}$ (Floridi, 2025a, “Formalising the Conjecture”; Floridi, 2025b, “Formalising the Conjecture”).

III. **Mapping Scope**: $\mathsf{S}(\mathsf{M}) \coloneqq \mathsf{K}(\mathrm{I}) + \mathsf{K}(\mathrm{O})$, with $\mathsf{K}(\cdot)$ the Kolmogorov complexity.

**Note 2.2:** Scope is defined as the sum of the Kolmogorov complexity of the input and output domains as stated in §2 (I). However, the published conjecture does not specify the $\mathsf{K}(\cdot)$ variant and the technical requirement to fix an effective *injective encoding* of those spaces (see §2.1(A), §2.1(B), and §2.1(C)).

## 2.1. Conventions and Standing Assumptions

The published conjecture does not specify whether the Kolmogorov complexity $\mathsf{K}(\cdot)$ is prefix-free or plain/classical, relative to a universal Turing machine (UTM), or an admissible encoding for $\mathrm{I}, \mathrm{O}$ sets. These choices are standard prerequisites in AIT; for this reason, without changing any published primitives, we set the minimal AIT framework below to make the conjecture formally testable in our disproof (see §§3.1–3.3). Additionally, given the lack of a formal definition of ‘sufficiently expressive mechanisms’, we introduce a conservative test class (*admissible mechanisms*; §2.1(E)) to render the conjecture formally testable.

A) **Universal machine.** Fix an optimal UTM $\mathsf{U}$ of the appropriate type for each K-variant (prefix-free in §3.1; plain in §3.2); by invariance, choices affect $\mathsf{K}(\cdot)$ only up to $O(1)$.

B) **Encoding**. Fix the binary alphabet $\{0, 1\}$. Let $\{0, 1\}^+$ denote the set of non-empty binary strings. Fix once and for all a single admissible encoding $\langle\cdot\rangle$ from finite mathematical objects (e.g., finite strings over a fixed finite alphabet, finite tuples, and finite sets) (Li & Vitányi, 2019, Section 1.11) to $\{0, 1\}^+$ (Li & Vitányi, 2019, Section 1.4); $\langle\cdot\rangle$ is *total*, *computable*, and *injective*.

Therefore, we make explicit the scope as

$$\mathsf{S(M)} \coloneqq \mathsf{K}(\langle \mathsf{I} \rangle) + \mathsf{K}(\langle \mathsf{O} \rangle),$$

and we adopt this definition from now on in the present rebuttal.

C) **Variants of K**. Each branch of the disproof specifies the variant of $\mathsf{K}(\cdot)$: §3.1 uses Prefix-K; §3.2 uses Plain-K.

D) **Invariance Theorem.** Changing the UTM $\mathsf{U}$ (or composing $\langle\cdot\rangle$ with a total computable bijection on $\{0, 1\}^+$) alters $\mathsf{K}(\cdot)$ by at most an additive $O(1)$ (for Prefix-K (Li & Vitányi, 2019, Section 3.1), for Plain-K (Li & Vitányi, 2019, Section 2.1)).

E) **Admissible mechanisms and "sufficiently expressive mechanisms."**

The original article states the conjecture for 'sufficiently expressive' AI mechanisms without specifying a formal domain of quantification. To make the claim formally testable, we introduce a conservative test class that captures the minimal structure presupposed by the conjecture and by its motivating examples: finite (yet arbitrarily rich) domains, a total specification, and a total input–output mapping.

**Definition (admissible mechanism).** A mechanism $\mathsf{M}$ is admissible if $\mathsf{I}$, $\mathsf{O}$ (input and output) are finite and non-empty sets, and $\mathsf{M}$ induces a total mapping $f_{\mathsf{M}}\colon \mathsf{I} \to \mathsf{O}$. With finite $\mathsf{I}$, $\mathsf{O}$ and a total specification $\mathsf{Spec}\colon \mathsf{I} \to \mathsf{O}$, such an $f_{\mathsf{M}}$ is computable; in particular, the mechanism with $f_{\mathsf{M}}(x) \equiv \mathsf{Spec}(x)$ for all $x \in \mathsf{I}$ is admissible.

We assume no a priori upper bound on the descriptive richness of admissible mechanisms: the class includes mechanisms whose $\mathsf{I}$, $\mathsf{O}$ descriptions have arbitrarily large Kolmogorov complexity, so that $\mathsf{K}(\langle \mathsf{I} \rangle) + \mathsf{K}(\langle \mathsf{O} \rangle)$ (as defined in §2.1(B)) can grow without bound. This is not an additional primitive: it simply states that the admissible class is not artificially capped by an external complexity budget, consistently with the conjecture's own motivating settings (finite-domain symbolic devices, large but finite vocabularies). Accordingly, the disproof below is carried out within this admissible class. This refutes the conjecture under any operational interpretation of 'sufficiently expressive' that includes such minimal finite-domain mechanisms (or, more strongly, the subclass with unbounded description complexity). Conversely, any restriction of 'sufficiently expressive' that excludes this conservative test class would render the conjecture inapplicable to the finite-domain settings motivating its formal statement, undermining its intended role as a general computable and auditable bound.

F) **Quantifier structure (universality claim).**

For clarity, we fix the conjecture's logical form as the universal claim:

$$\exists \mathsf{k} \in (0,1)\ \forall \mathsf{M} \in \mathcal{M}\colon \mathsf{C(M)} \cdot \mathsf{S(M)} \leq \mathsf{k},$$

where $\mathsf{S(M)} \coloneqq \mathsf{K}(\langle \mathsf{I} \rangle) + \mathsf{K}(\langle \mathsf{O} \rangle)$ (as fixed in §2.1(B)) and $\mathcal{M}$ denotes the class of "sufficiently expressive" mechanisms. To render the conjecture formally testable under its published primitives, we restrict attention to the admissible subclass of $\mathcal{M}$ specified in §2.1(E). Since the conjecture is universal over $\mathcal{M}$, exhibiting a single admissible counterexample $M \in \mathcal{M}$ with $\mathsf{C(M)} \cdot \mathsf{S(M)} > \mathsf{k}$ suffices to refute universality.

## 3. Conjecture's Disproof

This work relies on foundational results in prefix coding and algorithmic information theory: Kraft (1949), McMillan (1956), Kolmogorov (1965), and Chaitin (1966). Related foundational developments include Levin (1974) and Chaitin (1975).

For formal completeness, we split the disproof into two branches, as explained in §2 and §2.1:

(i) **Prefix-K (self-delimiting; prefix-free)** (Li & Vitányi, 2019, Chapter 3), developed in Section 3.1.

(ii) **Plain-K (classical)** (Li & Vitányi, 2019, Chapter 2), developed in Section 3.2.

Section 3.3 generalizes the two into the *main theorem.*

### 3.1. Prefix-K Branch: "High-Certainty Inconsistency"

Conventions as fixed in §2.1: in this branch, $\mathsf{K}(\cdot)$ denotes prefix (self-delimiting) Kolmogorov complexity. Let $\mathsf{U}$ be an optimal prefix-free UTM; $\mathsf{Dom(U)}$ is prefix-free (hence satisfies Kraft's inequality), and programs are finite binary strings.

**Lemma 1.1 (no length-0 program on a universal prefix machine).** Since the empty string $\varepsilon$ is a prefix of every binary string, if $\varepsilon \in \mathsf{Dom(U)}$ then prefix-freeness forces $\mathsf{Dom(U)} = \{\varepsilon\}$, contradicting universality, hence $\varepsilon \notin \mathsf{Dom(U)}$ and every valid program has length $\ell \geq 1$. ■

**Coding-theory view (Kraft–McMillan)** (Li & Vitányi, 2019, Section 1.11). For any prefix code $\mathcal{P} \subseteq \{0,1\}^{+}$, and program $p \in \mathcal{P}$,

$$\sum_{p \in \mathcal{P}} 2^{-|p|} \leq 1$$

A length-0 codeword contributes $2^0 = 1$, leaving no mass for any second codeword—again impossible for a universal machine.

**Corollary 1.1 (prefix floor).** By Lemma 1.1, for every non-empty binary string $x$, $\mathsf{K}(x) \geq 1$. Therefore, for non-empty admissible encodings $\langle \mathsf{I} \rangle$, $\langle \mathsf{O} \rangle$, $\mathsf{S(M)} = \mathsf{K}(\langle \mathsf{I} \rangle) + \mathsf{K}(\langle \mathsf{O} \rangle) \geq 2$.

*Proof.* Since $\langle \mathsf{I} \rangle$ and $\langle \mathsf{O} \rangle$ are non-empty strings, Lemma 1.1 yields $\mathsf{K}(\langle \mathsf{I} \rangle) \geq 1$ and $\mathsf{K}(\langle \mathsf{O} \rangle) \geq 1$. Hence $\mathsf{S(M)} \geq 2$. ■

*Clarification*: The lower bound $K(x) \geq 1$ for every non-empty string $x$ follows from prefix-freeness of $\mathrm{Dom}(U)$ (and the impossibility of $\varepsilon \in \mathrm{Dom}(U)$ for any universal prefix machine), not from any idiosyncratic choice of $U$; hence no additive $O(1)$ change of reference machine can invalidate the fixed 1-bit floor.

**Proposition 1.1 (hyperbolic ceiling).** From $C(M) \cdot S(M) \leq k$ with $k \in (0, 1)$, and $C(M) \geq \frac{1}{2}$ it follows that $S(M) < 2$.

*Proof.* From $C(M) \cdot S(M) \leq k$ and $C(M) > 0$, divide both sides by $C(M)$ to obtain $S(M) \leq \frac{k}{C(M)}$. If $C(M) \geq \frac{1}{2}$, then $S(M) \leq 2 \cdot k < 2$ because $0<k<1$. ■

**Theorem 1.1 (inconsistency at high certainty).** For any $M$ with non-empty $I/O$, and $C(M) \geq \frac{1}{2}$, the *prefix floor* $S(M) \geq 2$ contradicts the *hyperbolic ceiling* $S(M) < 2$.

Moreover, if $C(M) = 1$ by provable-certainty regime corollary, then

$$S(M) \leq \frac{k}{C(M)} = k < 1$$

This is, a *fortiori,* incompatible with the 2-bit floor. ■

### 3.2. Plain-K Branch: "Existential Falsification"

Conventions as fixed in §2.1: in this branch, $K(\cdot)$ denotes plain Kolmogorov complexity relative to a fixed optimal (non-prefix) UTM $U$ (Li & Vitányi notation: $C_U(\cdot)$).

**Lemma 2.1 (unboundedness on an infinite computable image)** (Li & Vitányi, 2019, Section 2.3). Let $\boldsymbol{\mathcal{R}} \coloneqq \{\langle O \rangle \colon O \text{ a finite non-empty set}\} \subseteq \{0, 1\}^+$. Then $\boldsymbol{\mathcal{R}}$ is infinite and *computably enumerable* (c.e.). Moreover, $\forall\, T \in \mathbb{N}, \exists\, r \in \boldsymbol{\mathcal{R}}$ with $K(r) > T$.

*Proof.* There are infinitely many finite sets; since $\langle \cdot \rangle$ is total, injective, and computable, $\boldsymbol{\mathcal{R}}$ is an infinite c.e. set of strings. For plain complexity, for each fixed $T$, the set $\{x \colon K(x) \leq T\}$ is finite and in fact $|\{x \colon K(x) \leq T\}| \leq \sum_{i=1}^{T} 2^i = 2^{(T+1)} - 2$ (since the reference UTM $U$ is taken to be a partial function $U \colon \{0, 1\}^+ \rightharpoonup \{0, 1\}^+$, each program either diverges or halts with a single finite output string). Hence, some $r \in \boldsymbol{\mathcal{R}}$ satisfies $K(r) > T$. ■

**Theorem 2.1 (existential counterexample under plain-K).** For all $0 < k < 1$ and for all $C^* \in (0, 1]$, there exists $M_{ex}$ with $C(M_{ex}) \geq C^*$ and $S(M_{ex}) > \frac{k}{C^*}$, hence $C(M_{ex}) \cdot S(M_{ex}) > k$.

*Proof.* Choose any non-empty finite input set I and let $c_I \coloneqq K(\langle I \rangle)$.

Set the threshold $T^* \coloneqq \frac{k}{C^*} - c_I$, and define $B \coloneqq \max\{0, T^*\}$ with $B \in \mathbb{R}$.

By Lemma 2.1, pick $\mathrm{r} \in \boldsymbol{\mathcal{R}}$ with $\mathrm{K}(\mathrm{r}) \geq \lfloor \mathrm{B} \rfloor + 1$.

By definition of $\boldsymbol{\mathcal{R}}$, there exists a finite non-empty output set $\mathrm{O}$ such that $\langle \mathrm{O} \rangle = \mathrm{r}$; fix such an $\mathrm{O}$.

Fix any total specification $\mathsf{Spec}$ over $\mathrm{I}, \mathrm{O}$ sets ($\mathrm{I} \to \mathrm{O}$) and define $f_{\mathrm{Mex}} \equiv \mathsf{Spec}$. Since $\mathrm{I}, \mathrm{O}$ are finite and $\mathsf{Spec}$ is total, $f_{\mathrm{Mex}}$ is computable; thus, $\mathrm{M}_{\mathrm{ex}}$ is admissible under the published primitives[1].

Then $\sup_{x \in \mathrm{I}} \Pr[f_{\mathrm{Mex}}(x) \neq \mathrm{Spec}(x)] = 0$, so $\mathrm{C}(\mathrm{M}_{\mathrm{ex}}) = 1 \geq \mathrm{C}^*$.

Moreover, $\mathrm{S}(\mathrm{M}_{\mathrm{ex}}) = \mathrm{K}(\langle \mathrm{I} \rangle) + \mathrm{K}(\langle \mathrm{O} \rangle) = \mathrm{c}_{\mathrm{I}} + \mathrm{K}(\mathrm{r})$.

*Case check*:

(i) If $\mathrm{T}^* > 0 \Leftrightarrow \mathrm{B} = \mathrm{T}^*$, then $\lfloor \mathrm{B} \rfloor + 1 > \mathrm{B}$, so $\mathrm{K}(\mathrm{r}) \geq \lfloor \mathrm{B} \rfloor + 1 > \mathrm{T}^*$, and

$\mathrm{S}(\mathrm{M}_{\mathrm{ex}}) = \mathrm{c}_{\mathrm{I}} + \mathrm{K}(\mathrm{r}) > \mathrm{c}_{\mathrm{I}} + \mathrm{T}^* = \frac{k}{C^*}$.

(ii) If $\mathrm{T}^* \leq 0 \Leftrightarrow \mathrm{B} = 0$ and $\frac{k}{C^*} \leq \mathrm{c}_{\mathrm{I}}$, since $\mathrm{K}(\mathrm{r}) \geq \lfloor \mathrm{B} \rfloor + 1 = 1$,

hence $\mathrm{S}(\mathrm{M}_{\mathrm{ex}}) = \mathrm{c}_{\mathrm{I}} + \mathrm{K}(\mathrm{r}) > \mathrm{c}_{\mathrm{I}} \geq \frac{k}{C^*}$.

Therefore, since $\mathrm{C}(\mathrm{M}_{\mathrm{ex}}) = 1$, $\mathrm{C}(\mathrm{M}_{\mathrm{ex}}) \cdot \mathrm{S}(\mathrm{M}_{\mathrm{ex}}) = \mathrm{S}(\mathrm{M}_{\mathrm{ex}}) > \mathrm{k}$. ■

---

[1] **Implementation irrelevance.** Nothing in the published primitives requires $f_{\mathrm{M}}$ to be implemented as a lookup table: because $\mathrm{I}$ and $\mathrm{O}$ are finite and $\mathsf{Spec}\colon \mathrm{I} \to \mathrm{O}$ is total, there exists a finite program that implements $\mathsf{Spec}$ (e.g., by hard-coding a finite procedure), hence the mechanism can be realized algorithmically. Any attempt to exclude such perfect finite-domain mechanisms as “not sufficiently expressive” would also exclude the conjecture’s own motivating finite-domain regimes, severing the universality claim from its stated application domain.

### 3.3. Main Theorem: "No Universal Hyperbolic Bound for Kolmogorov-Complexity–Based Scope and Accuracy Functionals"

From Theorems 1.1 and 2.1 (§§3.1–3.2), we derive a general form of the result, stated as the *main theorem*.

**Accuracy Functional $\mathsf{A(M)}$ Axioms.**

Assume the following minimal axioms for an *accuracy functional*[2] $\mathsf{A(M)}$:

(i) *Range*: $\mathsf{A(M)} \in (0, 1]$.

(ii) *Perfect-mechanism normalization*: if $f_{\mathsf{M}}(x)$ matches $\mathsf{Spec}(x)$ on $\mathsf{I}$, $\forall\, x \in \mathsf{I}$, then $\mathsf{A(M)} = 1$.

(iii) *Compatibility*: taking $\mathsf{A(M)} \equiv \mathsf{C(M)}$ recovers certainty as fixed in §2 and used in §§3.1–3.2.

No additional properties (e.g., monotonicity, continuity, calibration) are required.

**Main Theorem (No Universal Hyperbolic Bound for Kolmogorov-Complexity–Based Scope and Accuracy Functionals).** Fix a universal Turing machine (UTM) $\mathsf{U}$ and a single admissible encoding $\langle\cdot\rangle$ (*total, computable, injective*). Let $\mathsf{S(M)} \coloneqq \mathsf{K}(\langle \mathsf{I} \rangle) + \mathsf{K}(\langle \mathsf{O} \rangle)$ with $\mathsf{K}(\cdot)$ instantiated once—either as prefix-free or plain Kolmogorov complexity—relative to $\mathsf{U}$. Let $\mathsf{A(M)}$ be any accuracy functional satisfying axioms (i)–(iii), then there exists no $\mathsf{k} \in (0,\ 1)$ such that $\mathsf{A(M)} \cdot \mathsf{S(M)} \le \mathsf{k}$ holds for all admissible mechanisms $\mathsf{M}$, under the conjecture's published primitives.

*Proof.* Immediate from Theorems 1.1 and 2.1 (§§3.1–3.2): by (iii) set $\mathsf{A} \equiv \mathsf{C}$; by (ii) choose an admissible $\mathsf{M}$ with $f_{\mathsf{M}} \equiv \mathsf{Spec}$ so $\mathsf{A(M)} = 1$. The respective scope bounds then contradict $\mathsf{S(M)} \cdot \mathsf{A(M)} \le \mathsf{k}$ for any fixed $\mathsf{k} \in (0, 1)$: Prefix-K: $\mathsf{S(M)} \ge 2$ while $\mathsf{A(M)} = 1 \Rightarrow \mathsf{S(M)} \cdot \mathsf{A(M)} \ge 2 > \mathsf{k}$; Plain-K: there exists $\mathsf{M}_{\mathsf{ex}}$ with $\mathsf{A}(\mathsf{M}_{\mathsf{ex}}) = 1$ and $\mathsf{S}(\mathsf{M}_{\mathsf{ex}}) > \mathsf{k}$. ∎

[2] **Admissible machine's *accuracy functionals*.** Parameters such as an *input distribution, error tolerance, robustness radius, divergence ball, aggregation scheme, or resource budget* may be integrated into the definition of $\mathsf{A(M)}$; the *main theorem* remains agnostic to these preferences provided that the axioms are satisfied.

## 4. Robustness

The robustness notes in this section apply to §§3.1–3.3 and, where relevant, to Appendix A, without altering or overlapping what is fixed in §2.1. They set out the common ground across the two proof routes and their synthesis, clarify which choices are merely representational and therefore do not affect the conclusions, and indicate how to read the results with consistent units and terminology.

(1) **Independence.** The prefix inconsistency and the plain counterexample are independent; either refutes the conjecture's generality.

(2) **Invariance.** By the invariance theorem (recalled in §2.1(D)), changing the reference machine or applying any effective recoding changes $K(\cdot)$ by at most an additive $O(1)$. Therefore, invariance cannot (i) eliminate the fixed Prefix-K floor used in Theorem 1.1 (§3.1), nor (ii) turn unbounded growth (even up to an additive constant) into a bounded quantity. Hence, the disproof is machine- and encoding-invariant up to $O(1)$.

(3) **Admissibility.** Constructions use total maps on finite, non-empty $I/O$ sets, and the single encoding $\langle\cdot\rangle$ fixed in §2.1. Any restriction that excludes finite, non-empty $I/O$ mechanisms would exclude canonical motivating cases (e.g., symbolic calculators, finite-state controllers, and modern large language models (LLM) deployments with finite vocabularies and finite input/output alphabets). Such an exclusion would therefore sever the conjecture from its stated application domain.

(4) **Well-posedness (units).** $S(M)$ is measured in bits—whether defined via $K(\cdot)$ in §§3.1–3.3 (description length in bits) or via Shannon entropy in Appendix A (bits per draw, and hence bits per sample under the standard normalization)—while $k$ is dimensionless; absent an explicit normalization (or unit conversion), the product bound is therefore dimensionally ambiguous as stated. This is a well-posedness issue of the conjecture's formulation, not a technical limitation of the disproof: our refutation does not depend on any particular normalization of $S(M)$, and survives any fixed rescaling (which merely rescales $k$).

(5) **Robustness to scope revision.** Appendix A shows that an entropy-based revision (replacing Kolmogorov scope with Shannon joint entropy and redefining certainty accordingly) still cannot yield a universal constant: unless protocols are explicitly restricted, the product form remains unbounded across admissible mechanisms.

## 5. Immediate Implications

This section briefly distills two immediate implications of our formal results for the motivating regimes discussed in the original conjecture itself (Floridi, 2025a, "Some Logical Implications"; Floridi, 2025b, "Some Logical Implications"). In doing so, it also flags a deeper conceptual—potentially ontological—issue in the framing of the "certainty–scope" relationship, without attempting to settle it here. This section does not attempt to cover the broader discussion or the full range of philosophical, technical, and governance implications. The present work is, by design, a logical–mathematical disproof; the concepts introduced herein may be further extended and developed in dedicated work.

### 5.1. Risks for Wide-Scope Regimes (Subsymbolic AI/GenAI)

As currently formulated, the bound imposes a decrease in admissible certainty values as a function of increasing scope, independently of any additional evidential structure beyond the scalar product constraint itself. More importantly, it does not provide a range of admissible certainty values, which risks leading to misleading conclusions for large-scale AI systems.

### 5.2. Collapse in High-Certainty Regimes (Symbolic AI)

For $C(M) \geq 1/2$, the bound collapses into a sub-two-bit $S(M)$ ceiling, incompatible with standard self-delimiting coding facts; thus, the conjecture trivializes scope in the very regime used to motivate a high-certainty core.

## 6. Conclusion

As stated in the *main theorem*, no universal "certainty-scope" hyperbola exists. Under the conjecture's published primitives, there is no constant $k \in (0, 1)$ such that $C(M) \cdot S(M) \leq k$ holds for all admissible mechanisms. The Prefix-K branch yields a high-certainty inconsistency ($S(M) \geq 2$ vs. $S(M) < 2$), while the Plain-K branch provides an existential counterexample, $M_{ex}$, with arbitrarily wide scope at arbitrarily high certainty. Thus, from a logical–mathematical standpoint, the "certainty–scope" conjecture, in the form stated under its own primitives, is false. Moreover, Appendix A shows that a subsequent "entropy-based" revision cannot restore universality: even after replacing Kolmogorov scope with Shannon joint entropy and redefining epistemic certainty accordingly, no single constant bounds the product form across admissible mechanisms.

---

## Appendix A: The Entropy-Scope Revision (Shannon Joint Entropy) Also Cannot Restore Universality

### A.1 Motivation and Scope

A recent proposal, published in *Philosophy & Technology* (Messina, 2026), advances a revision of the conjecture by (i) replacing the original notion of *epistemic certainty* with a user-centric expected correctness/acceptability level (Epistemic Certainty Level, ECL), and (ii) substituting the Kolmogorov scope with Shannon joint entropy (Shannon, 1948), while retaining a universal constant $\mathsf{k}$ independent of the specific AI mechanisms $\mathsf{M}$.

More precisely, this revision defines the mapping scope as the joint Shannon entropy of the input–output pair under a deployment distribution, and replaces worst-case correctness with an ECL ($\mathsf{C}(\mathsf{M})$, following the original notation) given by an expectation of user-dependent and inference-dependent acceptability indicators, while still asserting the existence of a positive constant, independent of any specific system, for all 'sufficiently expressive' AI mechanisms.

This appendix shows that the universality claim remains false under this entropy-based scope: a counterexample follows immediately from elementary Shannon identities and from deterministic special cases explicitly covered by the revision proposal. Since the revised conjecture is formulated to hold for every sufficiently expressive AI mechanism $\mathsf{M}$, and in our framework, 'sufficiently expressive' is conservatively instantiated by the class of *admissible mechanisms* introduced in §2.1(E), a single admissible mechanism that violates the inequality already falsifies the claimed universality.

### A.2 Definitions (Entropy-based Scope)

Let $\mathsf{X}$ and $\mathsf{Y}$ be random variables taking values in finite sets $\mathsf{I}$ (input) and $\mathsf{O}$ (output), respectively, with joint distribution $P_{\mathsf{X,Y}}$ induced by the evaluation protocol. Define the *entropy-based scope* as:

$$\mathsf{S_H}(\mathsf{M}) := \mathsf{H}(\mathsf{I},\mathsf{O}) = -\sum_{(x,y)\in \mathsf{I}\times\mathsf{O}} P_{\mathsf{X,Y}}(x,y)\log_2 P_{\mathsf{X,Y}}(x,y)$$

*Notation*: For brevity, we write $\mathsf{H}(\mathsf{I},\mathsf{O}) \equiv \mathsf{H}(\mathsf{X},\mathsf{Y})$.

The revised universality claim has the same product form (Floridi, 2025a): there exists a constant $\mathsf{k}>0$, asserted to be independent of any specific mechanism $\mathsf{M}$, such that $\mathsf{A}(\mathsf{M})\cdot \mathsf{S_H}(\mathsf{M}) \le \mathsf{k}$ for all admissible $\mathsf{M}$, where $\mathsf{A}(\mathsf{M})$ satisfies the minimal axioms in §3.3. In particular, $\mathsf{A}(\mathsf{M})$ may be instantiated as ECL; the proof below uses only perfect-mechanism normalization (i.e., existence of admissible mechanisms with $\mathsf{A}(\mathsf{M}) = 1$).

### A.3 Lemmas

**Lemma A.1 (Deterministic outputs collapse joint entropy).** If outputs are deterministic given inputs (i.e., $Y=g(X)$ almost surely under $P_{X,Y}$), then $H(I, O) = H(I) + H(O|I) = H(I)$.

*Proof.* By the entropy chain rule (Cover & Thomas, 2006), $H(X, Y) = H(X) + H(Y|X)$. If $g(X) = Y$ (i.e., $Y$ is a deterministic function of $X$), then $H(Y|X) = 0$; hence, $H(X, Y) = H(X)$.■

**Lemma A.2 (Uniform inputs achieve $\log_2|I|$).** If $X \sim Unif(I)$, then $H(I) = \log_2|I|$.

*Proof.* If $X \sim Unif(I)$, then $P(X = x) = \frac{1}{|I|}$, for all $x \in I$. Thus $H(X) = -\sum_{x \in I} \frac{1}{|I|} \log_2\left(\frac{1}{|I|}\right) = \log_2|I|$. (see also Cover & Thomas, 2006).■

**Lemma A.3 (Perfect-mechanism normalization).** If $f_M(x)$ matches $Spec(x)$ on I for all $x \in I$, then $A(M) = 1$.

*Proof.* This is axiom (ii) in §3.3.■

**Note A.1**: This normalization is consistent with the 'Deterministic oracles' row of the summary table (Messina, 2026; p. 10), which explicitly assigns $C(M) = 1$ and scope $H(I)$ to a perfect deterministic mechanism, confirming that the counterexample operates within the revised conjecture's own scope of application.

**Note A.2:** Under the same protocol-induced input distribution (and induced joint distribution $P$ on $I \times O$), the chain rule gives $H(I, O) = H(I) + H(O|I) \geq H(I)$, since conditional entropy is non-negative. Thus, introducing stochasticity in the output cannot decrease the entropy-based scope relative to the deterministic case.

### A.4 Main Counterexample Theorem

**Theorem A.4 (No universal constant k holds under entropy-based scope).** There exists no constant $k>0$, independent of $M$, such that $A(M) \cdot S_H(M) \leq k$ holds for all admissible mechanisms $M$ under the entropy-based scope $S_H(M) = H(I, O)$.

*Proof.* Let $k>0$ be arbitrary. Choose $n = \lceil k \rceil + 1$ and take a finite input set I with $|I| = 2^n$. Fix any total specification $Spec: I \to O$ and consider the admissible perfect mechanism $M$ with

$f_M \equiv Spec$. By Lemma A.3, $A(M) = 1$. Under a deterministic specification, outputs are deterministic given inputs, so Lemma A.1 yields $S_H(M) = H(I, O) = H(I)$. Choose an evaluation protocol that induces $X \sim Unif(I)$; then, by Lemma A.2, $H(I) = \log_2|I| = n$.

Therefore, $A(M) \cdot S_H(M) = 1 \cdot n > k$ contradicting the claimed universal bound.■

Since $k>0$ is arbitrary, for every candidate $k$ there exists an admissible mechanism $M$ (under an admissible protocol) such that $A(M) \cdot S_H(M) > k$; hence, no universal constant can bound the product under entropy-based scope.

**Note A.3**: The revised conjecture asserts a universal constant $k$ while placing no explicit restriction on the admissible evaluation protocol class (e.g., input distributions or evaluation procedures). Accordingly, the universality claim is tested over admissible $\{M, \text{protocol}\}$ pairs, and a single admissible counterexample suffices to refute it. Any later restriction to “realistic” protocols or architectures would therefore be a substantive reformulation of the published claim, not a clarification of it.

### A.5 Implication: Any “Salvage” Breaks Universality

Theorem A.4 shows that substituting $K(\cdot)$ with $H(\cdot)$ cannot restore universality. This conclusion is relative to the conjecture’s published scope primitives, namely scope computed solely from I/O-level quantities (Immediato, 2025): in the original formulation via $S(M) \coloneqq K(\langle I \rangle) + K(\langle O \rangle)$, and in the entropy-based revision via an I/O-induced joint entropy term $H(X, Y)$. In this case, any attempt to salvage an inequality of the same product form must therefore impose a non-trivial external restriction, e.g., a uniform bound $H(X, Y) \leq W$ or a uniform bound on input support $|\text{supp}(X)| < N$. Consequently, any bound necessarily becomes parameterized by that external constraint: one can at best obtain statements of the form

$$A(M) \cdot S_H(M) \leq k(W) \text{ or } A(M) \cdot S_H(M) \leq k(N),$$

where the “constant” depends on a chosen entropy or support bound. This is incompatible with the revised conjecture’s claim that there exists a single $k>0$, ‘independent of any specific system $M$,’ that holds uniformly across all AI mechanisms and deployment protocols. A further external assumption—shared by both the original and the revised conjecture—is that an operational specification is available (either a *ground-truth* $\text{Spec}$ or a user-conditioned acceptability relation $\text{Spec}_u$), an assumption that becomes non-trivial precisely when symbolic and generative AI are treated as lying on a single deterministic-to-open-ended spectrum (Immediato, 2025).

Together with the *main theorem* (§3.3), Theorem A.4 establishes that the conjectured trade-off fails under both Kolmogorov–based and Shannon–entropy–based formulations.